\documentclass[superscriptaddress,showpacs,aps,prb,twocolumn]{revtex4}

\usepackage{times}
\usepackage{amsmath}
\usepackage{amssymb}
\usepackage{graphicx}
\usepackage{pst-all}

\begin{document}

\title{Coulomb Blockade Doppelg\"{a}ngers in Quantum Hall States}

\author{Parsa~Bonderson}
\affiliation{Microsoft Research, Station Q, Elings Hall, University of California, Santa Barbara, CA 93106, USA}

\author{Chetan~Nayak}
\affiliation{Microsoft Research, Station Q, Elings Hall, University of California, Santa Barbara, CA 93106, USA}
\affiliation{Department of Physics, University of California, Santa Barbara, CA 93106, USA}

\author{Kirill~Shtengel}
\affiliation{Department of Physics and Astronomy, University of California at Riverside, Riverside, CA 92507}
\affiliation{Institute for Quantum Information, California Institute of Technology, Pasadena, California 91125, USA}


\begin{abstract}
In this paper, we ask the question: How well can Coulomb blockade experiments correctly identify and distinguish between different topological orders in quantum Hall states? We definitively find the answer to be: Quite poorly. In particular, we write the general expression for the spacing of resonance peaks in a simple form that explicitly displays its dependence on the conformal scaling dimensions of the systems' edge modes. This form makes transparent the general argument that the Coulomb blockade peak spacings do not provide a strongly indicative signature of the topological order of the system, since it is only weakly related to the braiding statistics. We bolster this general argument with examples for all the most physically relevant non-Abelian candidate states, demonstrating that they have Coulomb blockade doppelg\"{a}ngers -- candidate states at the same filling fraction with identical Coulomb blockade signatures, but dramatically different topological orders and braiding statistics.
\end{abstract}

\date{\today}

\maketitle

\section{Introduction}

Quantum Hall states are remarkable physical systems because they are topologically ordered phases of matter. Non-Abelian topological phases, which possess quasiparticles whose exchange statistics are described by multi-dimensional representations of the braid group~\cite{Leinaas77,Goldin85,Fredenhagen89,Froehlich90}, have recently attracted much attention due to their potential use as a naturally fault-tolerant medium for quantum information processing~\cite{Kitaev03,Preskill98,Freedman98,Freedman03b,Preskill-lectures,Nayak08,Bonderson08a,Bonderson08b}. A number of non-Abelian quantum Hall states have been proposed as likely candidates for the second Landau level platueaus~\cite{Moore91,Blok92,Read99,Lee07,Levin07,Bonderson07d,Bishara08b}. The second Landau level quantum Hall states have been the focus of many recent experiments, including ones that have provided evidence in support of a non-Abelian state at $\nu=5/2$~\cite{Radu08,Willett09a}.

\begin{figure}[b!]
\begin{center}
  \includegraphics[scale=.65]{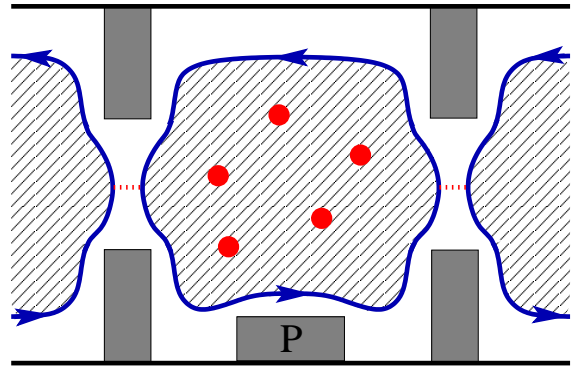}
  \caption{A quantum Hall liquid (hatched regions) with two constrictions generated by top gates (grey rectangles) which deplete the Hall liquid underneath them. This forms an isolated puddle with tunnel junctions on both sides, where electrons can tunnel (dashed lines) to and from the puddle, allowing some current to flow through the blockade. Tunneling resonance patterns may be observed in the edge current (arrows) as a result of changing the area of the puddle with a side gate $P$. These resonance patterns will depend on the total topological charge of the quasiparticles (red dots) contained in the puddle.}
  \label{fig:blockage}
\end{center}
\end{figure}

Clearly, it is important to have experimental tests that can accurately identify and distinguish the topological order physically realized in a quantum Hall system. A class of experiments proposed in Refs.~\onlinecite{Stern06a,Ilan08a} for such purposes is the Coulomb blockade experiments. In these experiments, a region of the quantum Hall liquid is effectively pinched off on two sides from the rest of the liquid, forming an isolated puddle. (Such a configuration may unintentionally occur when attempting to implement a double point-contact interferometer, such as those used for interference experiments. However, this appears not to be the case in the experiments of Ref.~\onlinecite{Willett09a}.) In such a configuration (see Fig.~\ref{fig:blockage}), most of the edge current will flow from one edge to the other at the pinched regions. However, some of the current can flow from one end of the Hall bar to the other as a result of electrons tunneling to and from the puddle, across the pinched off regions where the Hall liquid does not exist. Generically, such electron tunneling will be energetically prohibited by the charging energy of the puddle. However, there will be tunneling resonances when the ground-state energies are degenerate for two different values of $N_e$, the number of electrons in the pinched-off puddle. By varying the area of the puddle, one will pass through a series of such resonances, the spacing of which is determined by the properties of the particular quantum Hall state.

In this paper, we study the signatures of quantum Hall states in the Coulomb blockade experiments, demonstrating through general arguments and specific examples that this class of experiments does not provide a very discerning probe of topological order. Specifically, we write the general expression for the spacing between tunneling resonance peaks in a simple form which allows the predictions of this experimental signature to be obtained with only knowledge of the fusion rules and the conformal scaling dimensions of the edge modes of the candidate states. From this, we demonstrate that generally any quantum Hall state can have a Coulomb blockade doppelg\"anger -- a dramatically different quantum Hall state at the same filling fraction with identical Coulomb blockade signatures. This was first noticed while examining $\nu=5/2$ candidates in Ref.~\onlinecite{Bishara09}, where it was shown that the Abelian $(3,3,1)$ state~\cite{Halperin83} (without spin/component symmetry breaking) has the same Coulomb blockade pattern as the non-Abelian Moore-Read (MR)~\cite{Moore91}, anti-Pfaffian ($\overline{\text{Pf}}$)~\cite{Lee07,Levin07}, and SU$(2)_2$ non-Abelian fractional quantum Hall (NAF)~\cite{Blok92} states. In this paper, we examine all the strong candidate quantum Hall states and find doppelg\"{a}ngers for most of them. Furthermore, we show that nearly all of the viable candidates for the observed second Landau level filling fractions are Coulomb blockade doppelg\"{a}ngers of each other (at a given filling).

\section{Coulomb Blockade for Quantum Hall States}
\label{sec:CBGen}

In the general setting, Coulomb blockade experiments allow one to probe
correlations in a strongly-interacting system by the means of single
electron tunneling through a quantum dot via two tunnel junctions~\cite{Vanhouten92}. One
then expects to see resonant tunneling peaks, as a function of some
external parameters, whenever the energy of the isolated dot with $N$ and
$N+1$ electrons is degenerate, $\mathcal{E}(N)=\mathcal{E}(N+1)$. In the quantum Hall regime, one can envision several experimental settings. In
general, there may be several edge channels surrounding incompressible
regions at different filling fractions. The aforementioned ``dot'' need not
be fully isolated since not all edge channels are necessarily pinched off at
the tunnel junctions. Furthermore, the pinched-off region may itself
contain compressible regions~\cite{Rosenow07}.

To avoid ambiguity, we focus here on the case where the bulk of the
sample is in the quantum Hall state at a filling fraction $\nu$. The entire edge
structure then consists of $\lfloor{\nu}\rfloor$ outer integer quantum Hall channels
(separated by the incompressible strips) and the innermost fractional quantum Hall
edge corresponding to a $\tilde{\nu} \equiv \nu - \lfloor \nu \rfloor$ state. We
assume that only the innermost edge separating the incompressible
regions with filling fractions of $\lfloor{\nu}\rfloor$ and $\nu$ is
pinched off at the gated constrictions as shown in
Fig.~\ref{fig:blockage}. The integer quantum Hall channels that propagate past these
constrictions are not shown in the figure. We consider the case where the
pinched off puddle region contains incompressible Hall fluid. The puddle can contain a number of quasiparticle excitations, however the energy gap to their creation is assumed large comparing to the typical charging energies (this is just a restatement of the incompressible nature of the puddle). For the purpose of this treatment, we envision the experimentally tunable external parameters are the uniform background magnetic field and the voltage applied to the side gate $P$, which changes the equilibrium area of the puddle $A$.

The isolated nature of the puddle guarantees that it contains an integer
number of electrons -- this is a conventional Coulomb blockade setting. We
contrast this situation with the recently emerged notion of ``Coulomb
domination''~\cite{Rosenow07,Zhang09,Ofek09} whereby the Coulomb energy of
the puddle can be the dominant energy even when the puddle is far from
being pinched off from the rest of the Hall fluid. In the
Coulomb-dominated regime, the number of electrons in the puddle is
determined by the condition that they exactly neutralize the positively
charged background, but need not be quantized.

To translate the resonant tunneling condition $\mathcal{E}(N)=\mathcal{E}(N+1)$ (where the energy is predominantly Coulombic) to our setting, we notice that the
electron number can only change by one whenever the gate voltage is
increased by enough to allow one additional electron into or out of the
puddle. At this point, there is a peak in the longitudinal conductance
(which are also peaks in the longitudinal resistance, since
${R_L}\ll{R_H}$) since it is only at this point (or rather within ${k_B}T$ of it)
that the charge on the puddle can fluctuate. Since the density in the
puddle is fixed, the spacing between peaks as a function of area is
naively just the additional area required to allow one more electron into
the puddle:
\begin{equation}
\Delta A = \frac{e}{\rho_0}
\end{equation}
where $\rho_0 = e \tilde{\nu} B / \Phi_0$ is the charge density inside the dot. This consideration,
however, is too simplistic as it misses the quantum mechanical nature of
the edge. Specifically, the edge modes need to satisfy certain boundary
conditions that are consistent with their quantum numbers. In order to
incorporate this physics into our treatment, we follow
Ref.~\onlinecite{Ilan08a} and write the energy of a charged
edge as:
\begin{equation}
\mathcal{E}_c=\frac{v_c}{4\pi\tilde{\nu}} \int_{0}^{L} dx \left(\partial_x \varphi - 2\pi\tilde{\nu}
\frac{B (A-A_{0})}{L\Phi_0}\right)^2,
\end{equation}
where $L$ is the puddle circumference and $\Phi_0\equiv hc/e$ is the
magnetic flux quantum. Here $\varphi$ is the chiral bosonic field
describing the edge charge mode; the corresponding linear charge density
is given by $\rho=\partial_x \varphi/{2\pi}$. This description, however,
does not capture the entire edge physics of systems which have more than just such a charge mode. For general quantum
Hall states, one also needs to account for a kinetic energy of neutral edge modes. To do so, we turn to a more
formal description of excitations in the quantum Hall systems of interest.

Anyonic quasiparticles carry conserved quantum numbers called topological charge, which obey the fusion rules
\begin{equation}
a \times b = \sum_{c} N_{ab}^{c} c
\end{equation}
corresponding to the topological order of the state, where the fusion coefficients $N_{ab}^{c}$ specify the number of ways charge $a$ and $b$ can combine to produce charge $c$. If $a$ and $b$ are non-Abelian charges, then $\sum_{c} N_{ab}^{c} >1$. The bulk of the pinched-off puddle will have some definite total collective topological charge
\begin{equation}
a_{N_e} = a_{e}^{N_e} \times a_{\text{qps}}
,
\end{equation}
determined by $N_e$ together with the topological charge of an electron $a_{e}$, and the total topological charge $a_{\text{qps}}$ of the bulk quasiparticle excitations in the puddle. Since electrons are Abelian, $a_{N_e}$ is uniquely specified given a definite value of $a_{\text{qps}}$. If the bulk quasiparticles are all Abelian, then $a_{\text{qps}} = \prod_{j} a_j$ is also uniquely specified, where $j$ indexes the bulk quasiparticles and $a_j$ is the topological charge of the $j^{th}$ bulk quasiparticle. When the bulk quasiparticles are non-Abelian, there can be multiple fusion channels, and so we write $a_{\text{qps}} \in \prod_{j} a_j$ to indicate that $a_{\text{qps}}$ is one of the allowed fusion channels of the quasiparticles. In this case, $a_{\text{qps}}$ will still have a definite value, since the puddle is isolated. The entire puddle must have trivial total topological charge $0$, so, to compensate for the bulk topological charge $a_{N_e}$, the edge of the puddle carries the conjugate topological charge $\bar{a}_{N_e}$. This topological charge determines which sectors of edge excitations are allowed to occur, and hence the energy spectrum of the edge excitations. Thus, the pattern of tunneling resonance peaks is determined by the ground-state energy $\mathcal{E} \left( N_e , B, A, a_{N_e} \right)$ of the puddle~\cite{Ilan08a}, which depends on the number of electrons $N_e$ in the puddle, the background magnetic field $B$, the puddle area $A$, and the collective topological charge $a_{N_e}$ of the bulk.

The edge of a quantum Hall fluid can be described using conformal field theory (CFT)~\cite{DiFrancesco97}. For a pure CFT on a circle of length $L$, the energy of a $m^{th}$ level descendent of the primary field $\varphi$ is $\frac{2 \pi v}{L} \left( h_{\varphi} +m \right)$, where $v$ is the velocity and $h_{\varphi}$ is the conformal scaling dimension of $\varphi$. For a quantum Hall system, there can be multiple edge modes, and the topological charge $\bar{a}$ on the edge determines which primary field $\bar{a}^{(\beta)}$ of the $\beta^{th}$ mode is present. For this analysis, we are interested in the ground-state energies, and can ignore descendents (let $m=0$). Some of the edge modes may couple to other quantities that break their conformal symmetry. We can thus write the energy as the sum over effective energies from the edge modes
\begin{equation}
\mathcal{E} \left( N_e , B, A, a_{N_e} \right) = \sum_{\beta} \frac{2 \pi v_{\beta}}{L} \tilde{h}_{a}^{(\beta)}
\end{equation}
where $L$ is the length of the puddle's perimeter, $v_{\beta}$ is the velocity of the $\beta^{th}$ mode, and $\tilde{h}_{a}^{(\beta)}$ is the effective scaling dimension of the $\beta^{th}$ edge mode. The effective scaling dimensions include any modification of these modes that arise when the CFT couples to other quantities. When there is no modification of an edge mode, one simply has $\tilde{h}_{a}^{(\beta)} = h_{a}^{(\beta)}$, the conformal scaling dimension of $a^{(\beta)}$. For example, the charge sector's conformal dimension of an excitation with electric charge $e^{\ast}$ is given by
\begin{equation}
h_{a}^{(c)}  = \frac{\tilde{\nu} }{2 } \left[ a^{(c) } \right]^{2}= \frac{1 }{2 \tilde{\nu}} \left( \frac{e^{\ast}}{e} \right)^{2}
.
\end{equation}
However, we include the electrostatic area dependence in the energy of the charge mode (denoted $c$) by writing~\cite{Ilan08a}:
\begin{eqnarray}
\label{eq:htilde_h}
\tilde{h}_{a}^{(c)} &=& \left( \sqrt{ h_{a}^{(c)} } - \sqrt{\frac{\tilde{\nu}}{2}} \frac{ B \left(A - A_{0} \right) }{ \Phi_0} \right)^{2} \\
\label{eq:htilde_a}
&=& \frac{\tilde{\nu}}{2} \left( a^{(c)} - \frac{ B \left(A - A_{0} \right) }{ \Phi_0} \right)^{2}
,
\end{eqnarray}
where $A_0$ is the area of the puddle with just enough quasiholes fewer than the given configuration in order to have $a^{(c)}=0$ (see Refs.~\onlinecite{Ilan08a,Ilan08b} for more clarifying details), and only the fractional part of the filling $\tilde{\nu}$, enters the expression because the fully filled Landau levels are treated as inert.

The expression in Eqs.~(\ref{eq:htilde_h},\ref{eq:htilde_a}) can be written somewhat more transparently as
\begin{eqnarray}
\tilde{h}_{a}^{(c)} &=& \frac{\tilde{\nu}}{2} \left( N_{\phi}^{q} + S_0 - N_{\phi} \right)^{2} \\
N_{\phi}^{q} &=& \frac{ N_e }{\tilde{\nu}} + \sum_{j} a_{j}^{(c)} \\
N_{\phi} &=&  \frac{ B A }{ \Phi_0}
,
\end{eqnarray}
where $N_{\phi}^{q}$ is the quantized number of fluxes ascribed to the electrons (which have $a_{e}^{(c)}=1/\tilde{\nu}$ flux per electron) and bulk quasiparticles (the $j^{th}$ quasiparticle having $a_{j}^{(c)}$ fluxes), $S_0$ is a finite (not necessarily integer-valued) shift, and $N_{\phi}$ is the actual number of magnetic fluxes through the puddle. In this form, the energy is seen to be due to the discrepancy between the actual number of fluxes through the puddle and the flux quantization condition. Alternatively, we can translate flux to charge and think of this as the energy cost for violating charge neutrality. From this, we can see that without the inclusion of neutral modes the spacing between resonance peaks would simply be $\Delta A = e/\rho_0$, the average area that a single electron occupies. However, the internal structures of a quantum Hall state can give rise to deviations from this simple behavior through the neutral modes. In order to determine this deviant behavior, one must track how the topological charges of the neutral modes change as electrons are added to and removed from the puddle.

The spacing between two tunneling resonance peaks, for example the resonance when $\mathcal{E}(N-1)=\mathcal{E}(N)$ and the resonance when $\mathcal{E}(N)=\mathcal{E}(N+1)$, which has the corresponding sequence of total topological charge of the puddle:
\begin{equation}
\ldots \rightarrow a_{N-1} \rightarrow a_{N} \rightarrow a_{N+1} \rightarrow \ldots
\end{equation}
will be given by
\begin{eqnarray}
\label{eq:DA}
\Delta A_{N} &=& \frac{e}{\rho_0} \left( 1 + \sum_{\beta \neq c} \frac{\tilde{\nu} v_{\beta} }{v_c} \delta^{2} \tilde{h}_{a_{N}}^{(\beta)} \right) \\
\label{eq:delta2h}
\delta^{2} \tilde{h}_{a_{N}}^{(\beta)} &=& \tilde{h}_{a_{N +1}}^{(\beta)} + \tilde{h}_{a_{N-1}}^{(\beta)} -2\tilde{h}_{a_{N}}^{(\beta)}
\end{eqnarray}
assuming that $\tilde{h}^{(\beta)}$ do not depend on $A$ for $\beta \neq c$. Eqs.~(\ref{eq:DA},\ref{eq:delta2h}) are general, yet simple expressions which allow one to easily predict the Coulomb blockade spacings for candidate quantum Hall states. To compute the Coulomb blockade resonance peak patters from these equations, one only needs a cursory understanding of CFT. Specifically, one only needs to know the fusion rules, the conformal dimensions of the different topological sectors, and the topological charge carried by the electrons.

When there is a single neutral sector, e.g. when there is only one neutral mode or when the neutral modes have equilibrated into a single sector with common velocity $v_n$, this becomes
\begin{eqnarray}
\label{eq:DAs}
\Delta A_{N} &=& \frac{e}{\rho_0} \left( 1 + \frac{\tilde{\nu} v_{n} }{v_c} \delta^{2} \tilde{h}_{a_{N}}^{(n)} \right) \\
\label{eq:delta2hs}
\delta^{2} \tilde{h}_{a_{N}}^{(n)} &=& \tilde{h}_{a_{N +1}}^{(n)} + \tilde{h}_{a_{N-1}}^{(n)} -2\tilde{h}_{a_{N}}^{(n)}
\end{eqnarray}
In the following examples, we will always assume the simplest case where all the neutral modes of a state have equilibrated, and use these equations. However, one should keep in mind that the neutral modes may not be fully equilibrated, in which case one would still have to use Eqs.~(\ref{eq:DA},\ref{eq:delta2h}) with multiple neutral velocities.

We note that neutral mode velocities are typically expected to be small compared to the charge mode velocity. For example, the experimental studies of Refs.~\onlinecite{Wan08a,Hu09} found $v_{n} \sim 0.1 v_{c}$ for the $\nu=5/2$ state. Clearly, this would make it experimentally challenging to resolve the deviations of the resonance peak spacings from the trivial spacing $\Delta A_{N} = \frac{e}{\rho_0}$.

\subsection{Bulk-Edge Relaxation}

When there are non-trivial electrically neutral excitations in a quantum Hall state, there will generally be tunneling of such excitations between the edge and the bulk quasiparticles that will not change the bulk energy, but may lower the edge energy. If the area of the puddle is changed slowly compared to the tunneling rate of such neutral excitations (which is governed by the distance of bulk quasiparticles from the edge), then the spacing between consecutive resonance peaks will be modified because the total topological charge on the edge (and in the bulk) will change as a result of the tunneling event. We emphasize that while such bulk-edge relaxation generally can occur for non-Abelian quantum Hall states, it can also occur for certain Abelian states as well.

For the resulting spacing when bulk-edge relaxation occurs~\cite{Ilan08b}, one replaces $\delta^{2} \tilde{h}_{a_{N}}^{(\beta)}$ in Eq.~(\ref{eq:DA}) with
\begin{equation}
\label{eq:delta2h'}
\delta^{2} \tilde{h}_{a_{N} }^{ \prime (\beta)}  = \tilde{h}_{a_{N-1}}^{(\beta)} -\tilde{h}_{a_{N}^{\prime}}^{(\beta)} + \tilde{h}_{a_{N +1}^{\prime}}^{(\beta)} -\tilde{h}_{a_{N}}^{(\beta)}
,
\end{equation}
where the primed topological charges are the ones that result after electron tunneling before relaxation has occurred, and the unprimed topological charges are the ones that result after relaxation has occurred while the puddle area is being increased, i.e. when the topological charge advances through the sequence of relaxation and electron tunneling
\begin{equation}
\ldots \xrightarrow{\text{relax}} a_{N-1} \xrightarrow{e} a_{N}^{\prime} \xrightarrow{\text{relax}} a_{N} \xrightarrow{e} a_{N+1}^{\prime} \xrightarrow{\text{relax}} \ldots
\end{equation}
($N-1$ and $N+1$ would be interchanged if the puddle area were being decreased). When the area of the puddle is changed quickly compared to the neutral excitation bulk-edge tunneling rate, the spacing will simply look like Eq.~(\ref{eq:DA}). For intermediate time scales, the spacing between consecutive resonance peaks will be given by some smearing between Eqs.~(\ref{eq:delta2h}) and (\ref{eq:delta2h'}). Bulk-edge relaxation generally has the effect of decreasing the visibility of bunching in the resonance peak patterns.

\subsection{Changing Quasiparticle Content}

In the above analysis, it was assumed that the quasiparticle content of the puddle remains fixed while the area of the puddle is changed. This, of course, need not be the case. For example, a quasiparticle could be pinned at an impurity site which passes from the inside to the outside of the puddle as its boundary is moved in the process of changing its area. When the quasiparticle content of the puddle is changed in the course of the experiment, the observed Coulomb blockade resonance peak pattern will switch between patterns corresponding to different total topological charge $a_{\text{qps}}$ of the quasiparticles contained in the puddle (and hence of different total topological charge $a_{N_{e}}$ of the puddle).

\subsection{Multiple Electron Flavors}

When there are multiple ``flavors'' of electrons (e.g. in multi-layer or spin unpolarized systems), one generalizes the above discussion in the obvious way. Specifically, each flavor of electron has a particular topological charge assigned to it, which generally differs from one another. Consequently, the additional or removal of an electron of a particular flavor may be energetically preferred or disfavored for a given configuration of the system. Such an energetic preference must be taken into account when determining the sequence of topological charge as electrons are added to or removed from the puddle, and hence of the spacing between resonance peaks.

\subsection{Non-Uniform Filling}

When there is non-uniform filling, e.g. for hierarchical states exhibiting a layered edge structure, one must treat each region of given filling fraction separately. Consequently, one has a distinct copy of Eqs.~(\ref{eq:DA},\ref{eq:delta2h}) for each edge separating two regions of different filling, with $\tilde{\nu}$ now equal to the difference in filling fractions between these two regions. However, in such a scenario with multiple regions of different filling and multiple edges, there is \emph{a priori} no relation between the areas of the different regions, since these are non-universal properties that will depend on system details. Because of this, it is difficult to make meaningful predictions regarding the resulting tunneling resonance patterns that arise for states with non-uniform filling.

\section{Coulomb Blockade Doppelg\"{a}ngers}

At first, one might naively think that the spacing patterns of Coulomb blockade tunneling resonance peaks should be rather distinctive signatures of the topological order of the system, since they are determined by the corresponding fusion rules and conformal scaling dimensions, which are in fact highly distinctive properties. However, a bit more thought reveals the fallacies of this reasoning: first of all, it is only a very restricted set of fusion rules that apply in this setting, i.e. repeated fusion with the Abelian topological charge carried by electrons; secondly, the conformal dimensions do not enter the expression Eq.~(\ref{eq:DA}) in a simple way, but rather do so in the combinations given in Eqs.~(\ref{eq:delta2h},\ref{eq:delta2h'}). From this, it is clear that quantum Hall states with dramatically different topological orders and braiding statistics (or, more specifically, even with dramatically different fusion rules and conformal scaling dimensions) can nonetheless give rise to Coulomb blockade patterns that are identical.

This complication goes beyond the experimental challenges which may make different states difficult to distinguishable via Coulomb blockade experiments, such as insufficient resolution, the likely small values of $v_{\beta} / v_c$, and the thermal smearing of resonance peaks. While these experimental obstacles may in principle be overcome, the indistinguishability of Coulomb blockade doppelg\"{a}ngers is an inherent problem which cannot be surmounted within this class of experiments. Hence, in contrast to interference experiments, Coulomb blockade lacks the ability to unambiguously identify even the presence of non-Abelian statistics in a quantum Hall state.

\section{Examples}

Having established simple methods which allow us to easily compute the Coulomb blockade tunneling resonance peak patterns, we now turn to the important examples of candidate quantum Hall states.

\subsection{U$(1)$ sectors}

If a quantum Hall state includes an array of Abelian U$(1)$ sectors with coupling $K$-matrix, it is useful to separate them into the charge and neutral modes. This can be done by directly diagonalizing the $K$-matrix, or alternatively at the level of the flux vector. In this manner, for an excitation with U$(1)_{K}$ flux vector $\overrightarrow{l}$, one can write~\cite{Wen95}
\begin{eqnarray}
a^{(c)} &=& \frac{e^{\ast}}{\tilde{\nu} e} = \frac{\hat{t}_c \cdot K^{-1} \cdot \overrightarrow{l}}{\hat{t}_c \cdot K^{-1} \cdot \hat{t}_c} \\
\label{eq:hc}
h_{a}^{(c)} &=& \frac{\left[ a^{(c) } \right]^{2}}{2} \,\, \hat{t}_c \cdot K^{-1} \cdot \hat{t}_c \\
\overrightarrow{l}^{(n)} &=& \overrightarrow{l} - a^{(c) } \hat{t}_{c} \\
\label{eq:hn}
h_{a}^{(n)} &=& \frac{1}{2} \left| \overrightarrow{l}^{(n)} \cdot K^{-1} \cdot \overrightarrow{l}^{(n)} \right|
,
\end{eqnarray}
where $\hat{t}_{c}$ is the ``charge vector'' of the corresponding $K$-matrix (and $\hat{t}_c \cdot K^{-1} \cdot \overrightarrow{l}_{n}=0$). (Note: these equations can also be used for electrons by treating them as excitations with $e^{\ast} = e$, rather than their actual charge $-e$.) The fusion rules of topological charges in such U$(1)$ sectors is given by addition of the flux vectors. This can apply to hierarchical/composite fermion type states, as considered in the next section, as well as multi-component (``component'' can mean layer, flavor, spin, etc.) states~\cite{Wen92a}.

\subsection{Haldane-Halperin States}

The Haldane-Halperin (HH) states~\cite{Haldane83,Halperin84} are described by an array of U$(1)$s, and so can be analyzed using their corresponding $K$-matrices and charge vectors (see, e.g. Ref~\onlinecite{Wen92a}), as previously explained (for hierarchy states, one must remember to identify flux vectors under addition of the electrically neutral bosons in order to produce the smallest conformal dimension). Unfortunately, this is cumbersome to apply in complete generality. However, the subset of these states at the prominent filling fractions $\nu = \frac{n}{2np \pm 1}$, which also admits an equivalent composite fermion (CF) description~\cite{Jain89}, possesses additional symmetry which allows the edge theories to be described by $\text{SU}(n)_{\pm 1} \times \text{U}(1)$, where the $\text{U}(1)$ is purely the charge sector and $\text{SU}(n)_{\pm 1}$ is the neutral sector~\cite{Froehlich91}. This separation into charge and neutral sectors makes this subset of HH states very easy to analyze. The $\text{SU}(n)_{\pm 1}$ charges obey $\mathbb{Z}_{n}$ fusion rules
\begin{equation}
\Lambda_{j_1} \times \Lambda_{j_2} = \Lambda_{\left[j_1+j_2 \right]_{n}}
\end{equation}
where we define $\left[j \right]_{n} = j \text{mod} n$, and have conformal dimensions
\begin{equation}
h_{\Lambda_{j}} = \frac{j \left( n-j \right)}{2n}
.
\end{equation}
The electron carries $\text{SU}(n)_{\pm 1}$ charge $\Lambda_{1}$. Thus, as the number of electrons in the puddle increases (one at a time), the total $\text{SU}(n)_{\pm 1}$ charge of the puddle advances through the sequence
\begin{equation}
\ldots \rightarrow \Lambda_{\left[j \right]_{n} } \rightarrow \Lambda_{\left[ j+1 \right]_{n}} \rightarrow \ldots
\end{equation}
for which one immediately obtains
\begin{equation}
\delta^{2} h^{(n)}_{\Lambda_j} = \left\{
\begin{array}{lll}
1-\frac{1}{n}  & \text{ for } &  j = 0 \\
-\frac{1}{n}   & \text{ for } &  j \neq 0
\end{array}
\right.
.
\end{equation}
This bunches the resonance peaks into groups of $n$ peaks. This matches the result obtained in Ref.~\onlinecite{Cappelli09} through the use of annulus CFT partition functions. One can also check that the same results are obtained using $K$-matrix methods for the calculation~\footnote{When using $K$-matrix methods for hierarchy states, one must use identifications of flux vectors under addition/subtraction of the $n-1$ neutral boson flux vectors, i.e. $l \sim l + B_{i}$ where $i=2,\ldots,n$ and $\left(B_{i}\right)_{j}= K_{ij}$, in order to reduce $h_{a}^{(n)}$ to its minimal value.}. Relaxation does not occur for these states. The $\nu= 1/m$ Laughlin states~\cite{Laughlin83} correspond to $n=1$ and $p = \frac{m-1}{2}$, which gives no neutral mode, and hence trivially has $\Delta A = e/ \rho_{0}$.

\subsection{$k$-Component $[M+2;M]$ States and their Hierarchical Counterparts}
\label{sec:kcomponent}

The multi-component Abelian U$(1)_{K}$ states~\cite{Wen92a} with $K_{ij} = M + 2 \delta_{ij}$ (where $M$ is an integer) for $i,j \in \left\{ 1,\ldots,k \right\}$ and charge vector $\hat{t}_c = \left( 1 \ldots 1 \right)^{T}$ are $\nu = \frac{k}{kM+2}$ generalizations of Halperin's $(3,3,1)$ state~\cite{Halperin83} (which is the $k=2$, $M=1$ case). Using Eq.~(\ref{eq:hn}) with $\left[K^{-1}\right]_{ij} = \frac{-M}{2 \left(kM+2\right)} + \frac{1}{2} \delta_{ij}$, we find
\begin{equation}
h^{(n)}_{ \overrightarrow{l} } = \frac{1}{4} \sum_{j} l_{j}^{2} - \frac{1}{4k} \left( \sum_{j} l_{j} \right)^{2}
.
\end{equation}
The $j$th layer electron $e_j$ has corresponding flux vector $\overrightarrow{l}_{(e_{j})}$ with components $l_{(e_{j}) i} = K_{ij}$. We define the special flux vectors
\begin{equation}
\overrightarrow{a}_{(q,m)} = \overrightarrow{a}_{(q)} + \sum_{j=1}^{m} \overrightarrow{l}_{(e_{j})}
\end{equation}
where $\overrightarrow{a}_{(q)} = \left( 0 \ldots 0 1 \ldots 1 \right)^{T}$ has its first $q$ entries equal to $0$ and its last $k-q$ entries equal to $1$. It is straightforward to compute
\begin{equation}
\label{eq:hn31}
h^{(n)}_{ \overrightarrow{a}_{(q,m)}} = \left\{
\begin{array}{lll}
\frac{ q\left( k-q \right) + 4qm - 4 m^2 }{4k}                         & \text{ for }   &  0 \leq m \leq q \\
\frac{ q\left( k-q \right) + 4qm - 4 m^2 }{4k} - \frac{q}{2} +m         & \text{ for }   &  q \leq m \leq k
\end{array}
\right.
.
\end{equation}
At this point, it may still not be obvious why we are paying special attention to the flux vectors $\overrightarrow{a}_{(q,m)}$. We first note that $h^{(n)}$ satisfies the property
\begin{equation}
\label{eq:h_relation}
h^{(n)}_{ \overrightarrow{l} } = h^{(n)}_{ \overrightarrow{l} + \hat{t}_c }
\end{equation}
and that permuting the components of a flux vector leaves its conformal dimension unchanged. This makes it clear that the flux vectors for an arbitrary energetically preferred tunneling sequence is related to the flux vectors $\overrightarrow{a}_{(q,m)}$ through permutation of components and (multiple) addition/subtraction of $\hat{t}_c$. When a flux vector cannot be related to one of $\overrightarrow{a}_{(q,m)}$ through permutation and addition/subtraction of $\hat{t}_c$, it represents an energetically unfavored state. Such flux vectors will reach ones that can be related to $\overrightarrow{a}_{(q,m)}$ after a few electrons have tunneled. Thus, we see that for an arbitrary configuration of bulk quasiparticles, electron tunneling will give a sequence of flux vectors that is equivalent (in terms of $h^{(n)}$ values) to
\begin{equation}
\ldots \rightarrow \overrightarrow{a}_{(q,m)} \xrightarrow{e_{m+1}} \overrightarrow{a}_{(q,m+1)} \rightarrow \ldots
,
\end{equation}
and hence the Coulomb blockade resonance peak spacing is determined by Eq.~(\ref{eq:hn31}) by successively increasing $m$, which gives
\begin{equation}
\delta^{2} h^{(n)}_{\overrightarrow{a}_{(q,m)}} = \left\{
\begin{array}{lll}
2-\frac{2}{k}   &  \text{ for }  &  \left[ m \right]_{k} = q = 0    \\
1-\frac{2}{k}   &  \text{ for }  &  \left[ m \right]_{k} = 0,q \text{ when } q \neq 0 \\
-\frac{2}{k}    &  \text{ for }  &  \left[ m \right]_{k} \neq 0,q
\end{array}
\right.
\end{equation}
This produces bunching of the Coulomb blockade resonance peaks into alternating groups of $q$ and $k-q$ peaks. This is identical to the RR$_{k,M}$ Coulomb blockade pattern (see Section~\ref{sec:RR}, with $q$ here matching up with $2j$ there).

For concreteness, we explicitly consider the $(3,3,1)$ state. If there are an even number of quasiparticles in the puddle, then electron tunneling will give either of the two sequences (up to equivalences):
\begin{eqnarray}
&& \ldots \rightarrow \left(
\begin{array}{c}
0 \\
0
\end{array} \right) \xrightarrow{e_1} \left(
\begin{array}{c}
3 \\
1
\end{array}
\right) \xrightarrow{e_2} \left(
\begin{array}{c}
0 \\
0
\end{array}
\right) \rightarrow \ldots \\
&& \ldots \rightarrow \left(
\begin{array}{c}
0 \\
0
\end{array} \right) \xrightarrow{e_2} \left(
\begin{array}{c}
1 \\
3
\end{array}
\right) \xrightarrow{e_1} \left(
\begin{array}{c}
0 \\
0
\end{array}
\right) \rightarrow \ldots
\end{eqnarray}
i.e. either component electron can tunnel when the total topological charge is trivial, but if there is an imbalance between components, tunneling an electron of the deficient component will be energetically preferred. This gives alternation between $\delta^{2} h^{(n)} = \pm 1$.
If there are an odd number of quasiparticles in the puddle, then electron tunneling will give the sequence (up to equivalences)
\begin{equation}
\ldots \xrightarrow{e_2} \left(
\begin{array}{c}
0 \\
1
\end{array} \right) \xrightarrow{e_1} \left(
\begin{array}{c}
1 \\
0
\end{array}
\right) \xrightarrow{e_2} \left(
\begin{array}{c}
0 \\
1
\end{array}
\right) \xrightarrow{e_1} \ldots
\end{equation}
which has $\delta^{2} h^{(n)} = 0$.

The $k$-component $\left[ M+2 ; M \right]$ states allow relaxation by tunneling electrically neutral excitations (e.g. $\left(-1 1 0 \ldots 0 \right)$) between the edge and bulk, or alternatively (but equivalently in effect) by
tunneling flux from one component to another. The fully relaxed state has its flux spread as evenly between the components as possible, i.e. producing flux vectors that are related to $\overrightarrow{a}_{(q)}$ through permutation and addition/subtraction of $\hat{t}_c$. Adding an electron and then relaxing the system in this manner, one has a tunneling and relaxation sequence that is equivalent to
\begin{equation}
\ldots \xrightarrow{\text{relax}} \overrightarrow{a}_{(q)} \xrightarrow{e_{q+1}} \overrightarrow{a}_{(q,1)} \xrightarrow{\text{relax}} \overrightarrow{a}_{(\left[ q+2 \right]_k)} \xrightarrow{e_{q+3}} \ldots
,
\end{equation}
thus giving
\begin{equation}
\delta^{2} h^{\prime (n)}_{\overrightarrow{a}_{(q)}} = \left\{
\begin{array}{lll}
1-\frac{2}{k}  & \text{ for } &  q = 0, 1 \\
-\frac{2}{k}   & \text{ for } &  q \neq 0,1 \\
\end{array}
\right.
.
\end{equation}
When $k$ is even, this gives bunching of the resonance peaks into groups of $k/2$. When $k$ is odd, this gives bunching of the peaks into alternating groups of $\frac{k-1}{2}$ and $\frac{k+1}{2}$. This is also identical to the RR$_{k,M}$ Coulomb blockade pattern when relaxation occurs (see Section~\ref{sec:RR}).

One can also construct Abelian single-component ($k^{th}$ level) hierarchical counterparts of these $k$-component states, by starting from $\nu=\frac{1}{M+2}$ Laughlin states and condensing paired fundamental quasielectrons. Specifically, these states are described (in the hierarchical basis) by the charge vector $\hat{t}_{c} = (1 0 \ldots 0)^{T}$ and the $K$-matrix with non-zero elements: $K_{11}=M+2$, $K_{jj} = 4$, and $K_{j,j-1}=K_{j-1,j}= -2$ for $j=2,\ldots, k$. It is clear that these hierarchical states have Coulomb blockade patterns that are identical to that of their $k$-component counterparts (and, hence, also of the RR$_{k,M}$ states), since their $K$-matrices are related by SL$(k,\mathbb{Z})$ transformations.

The $k$-component $\left[ M+2;M \right]$ states, their hierarchical counterparts, and the RR$_{k,M}$ states provide a rather demonstrative example of Coulomb blockade doppelg\"{a}ngers. In particular, comparing the conformal dimensions in Eqs.~(\ref{eq:hn31}) and (\ref{eq:h_Pf}) reveals how different even the conformal dimensions can be for states that produce the same Coulomb blockade patterns. However, it is perhaps not so surprising that these states are doppelg\"{a}ngers of the RR states, given the deep connections known to exist between them~\cite{Ho95,Cappelli01,Regnault08}.

Finally, we note that the Coulomb blockade patterns computed above for the $k$-component $[M+2;M]$ states assumed that there is either no or only very weak (so as not to be noticeable) breaking of the multi-component symmetry of the state. Clearly, if this symmetry were broken, the different energies would be affected, resulting in different Coulomb blockade patterns. Such symmetry breaking would allow the $k$-component $\left[ M+2;M \right]$ and RR$_{k,M}$ states to potentially be distinguishable using Coulomb blockade. However, the counterpart of this symmetry for the hierarchical counterpart states will not be broken if the edges are fully equilibrated. We will see further examples of Coulomb blockade doppelg\"{a}ngers for which there is no such resolution from symmetry breaking.

\subsection{Bonderson-Slingerland States}

The Bonderson-Slingerland (BS) hierarchy states~\cite{Bonderson07d} generalize the hierarchical construction to apply to non-Abelian states. The simplest of these simply applies hierarchy to the charge sector, adding an array of U$(1)$s to the parent state, which can then be analyzed with the help of corresponding $K$-matrices and charge vectors.

Similar to the HH states, there is a subset of these states which admits a CF type description~\cite{Bonderson07d}, which will also have a mapping of the edge theories to a simple form partitioned into charge and neutral sectors. Specifically, these are the BS states at filling fractions $\nu = \frac{n}{\left(2p \mp 1 \right)n \pm 1}$ (denoted BS$_\nu$) which are built on the $\nu = \frac{1}{2p}$ MR parent states (including the two series $\nu = \frac{n}{n +1}$ and $\frac{n}{3n-1}$ built on $\nu=1/2$). Their edge theories can be mapped to $\text{Ising} \times \text{SU}(n)_{\pm 1} \times \text{U}(1)$, where the $\text{U}(1)$ is purely the charge sector and Ising and $\text{SU}(n)_{\pm 1}$ are the neutral sectors. The Ising charges' fusion rules are
\begin{equation}
\begin{array}{lll}
I \times I = I, \quad & I \times \psi = \psi, \quad & I \times \sigma = \sigma, \\
\psi \times \psi = I, \quad & \psi \times \sigma = \sigma, \quad & \sigma \times \sigma = I+\psi ,
\end{array}
\label{eq:Isingfusion}
\end{equation}
and they have conformal dimensions
\begin{equation}
h_I =0, \quad h_{\psi}=1/2, \quad h_{\sigma}=1/16
.
\end{equation}
Noting that the electron carries Ising$\times \text{SU}(n)_{\pm 1}$ charge $\left( \psi, \Lambda_1 \right)$, we see that if the puddle contains an even number of $\sigma$-type quasiparticles, electron tunneling will give the sequences of (neutral sector) Ising$\times \text{SU}(n)_{\pm 1}$ charge
\begin{equation}
\ldots \rightarrow \left( I, \Lambda_{\left[j \right]_{n}} \right)  \rightarrow \left( \psi, \Lambda_{\left[ j+1 \right]_n } \right) \rightarrow \left( I, \Lambda_{\left[ j+2 \right]_n} \right) \rightarrow \ldots
\end{equation}
for which
\begin{eqnarray}
\delta^{2} h^{(n)}_{\left( I,\Lambda_j \right)} &=& \left\{
\begin{array}{lll}
2-\frac{1}{n}  & \text{ for } &  j = 0 \\
1-\frac{1}{n}  & \text{ for } &  j \neq 0
\end{array}
\right. \\
\delta^{2} h^{(n)}_{\left( \psi,\Lambda_j \right)} &=& \left\{
\begin{array}{lll}
-\frac{1}{n}    & \text{ for } &  j=0 \\
-1-\frac{1}{n}  & \text{ for } &  j \neq 0
\end{array}
\right.
\end{eqnarray}
which is like the HH/CF states, but in addition to bunching into groups of $n$, there is further pairwise bunching resulting when the Ising charge switches between $I$ and $\psi$.

When the puddle contains an odd number of $\sigma$-type quasiparticles, electron tunneling will give the sequences of (neutral sector) Ising$\times \text{SU}(n)_{\pm 1}$ charge
\begin{equation}
\ldots \rightarrow \left( \sigma, \Lambda_{\left[j \right]_{n}} \right)  \rightarrow \left( \sigma, \Lambda_{\left[ j+1 \right]_n } \right) \rightarrow \ldots
\end{equation}
for which
\begin{equation}
\delta^{2} h^{(n)}_{\left( \sigma,\Lambda_j \right)} = \left\{
\begin{array}{lll}
1-\frac{1}{n}   & \text{ for }  &  j=0 \\
-\frac{1}{n}    & \text{ for }  &  j \neq 0
\end{array}
\right.
\end{equation}
which is exactly like the HH/CF states, without any additional pairwise bunching resulting from the Ising sector.

When $n$ is odd, these states allow relaxation by tunneling a neutral $\psi$ charge between the edge and bulk quasiparticles. Adding an electron to the puddle and then relaxing the system in this manner, one has the tunneling and relaxation sequence
\begin{equation}
\ldots \xrightarrow{\text{relax}} \left( I, \Lambda_{\left[j \right]_{n}} \right) \xrightarrow{e} \left( \psi, \Lambda_{\left[j+1 \right]_{n}} \right) \xrightarrow{\text{relax}} \left( I, \Lambda_{\left[j+1 \right]_{n}} \right) \xrightarrow{e} \ldots
\end{equation}
when the puddle contains an even number of $\sigma$-type quasiparticles, and
\begin{equation}
\ldots \xrightarrow{e} \left( \sigma, \Lambda_{\left[j \right]_{n}} \right)  \xrightarrow{e} \left( \sigma, \Lambda_{\left[ j+1 \right]_n } \right) \xrightarrow{e} \ldots
\end{equation}
(i.e. is unaffected by relaxation) when the puddle contains an odd number of $\sigma$-type quasiparticles. Both of these have
\begin{equation}
\delta^{2} h^{\prime (n)}_{\Lambda_j } = \left\{
\begin{array}{lll}
1-\frac{1}{n}  & \text{ for } &  j=0 \\
-\frac{1}{n}   & \text{ for } &  j \neq 0
\end{array}
\right.
\end{equation}
wherein the Coulomb blockade resonance peak spacing is independent of the bulk Ising charge of the puddle, and bunches peaks into groups of $n$.

When $n$ is even, these states allow relaxation by tunneling either a neutral $\psi$ charge or a neutral $(\sigma , \Lambda_{n/2})$ charge between the edge and bulk quasiparticles. For this we define $j_{\pm} = \left\lfloor \frac{2n \pm 1}{8} \right\rfloor$. (Note that $j_{+}+j_{-}+1 = n/2$.) Adding an electron to the puddle and then relaxing the system in this manner, the tunneling and relaxation sequence (which occurs for all bulk quasiparticle configurations) is
\begin{equation}
\ldots \xrightarrow{\text{relax}} \left( I, \Lambda_{\left[j \right]_{n}} \right) \xrightarrow{e} \left( \psi, \Lambda_{\left[j+1 \right]_{n}} \right) \xrightarrow{\text{relax}} \left( I, \Lambda_{\left[j+1 \right]_{n}} \right) \xrightarrow{e} \ldots
\end{equation}
for $0 \leq j < j_{+}$ and $n-j_{+} \leq j \leq n$,
\begin{equation}
\ldots \xrightarrow{e} \left( \sigma, \Lambda_{\left[j \right]_{n}} \right)  \xrightarrow{e} \left( \sigma, \Lambda_{\left[ j+1 \right]_n } \right) \xrightarrow{e} \ldots
\end{equation}
for $0 \leq j < j_{-}$ and $n-j_{-} \leq j \leq n$, and
\begin{equation}
\ldots \xrightarrow{\text{relax}} \left( I, \Lambda_{j_{+} } \right) \xrightarrow{e} \left( \psi, \Lambda_{j_{+}+1} \right) \xrightarrow{\text{relax}} \left( \sigma, \Lambda_{n-j_{-}} \right)  \xrightarrow{e} \ldots
\end{equation}
\begin{equation}
\ldots \xrightarrow{e} \left( \sigma, \Lambda_{j_{-}} \right) \xrightarrow{e} \left( \sigma, \Lambda_{j_{-}+1} \right) \xrightarrow{\text{relax}} \left( I, \Lambda_{n-j_{+}} \right) \xrightarrow{e} \ldots
\end{equation}
for $j=j_{\pm}$, respectively. For $n>4$, this gives
\begin{eqnarray}
\delta^{2} h^{\prime (n)}_{\left( I, \Lambda_{j} \right)} &=& \left\{
\begin{array}{lll}
1-\frac{1}{n}  & \text{ for } &  j=0 \\
-\frac{1}{n}   & \text{ for } &  j \neq 0,
\end{array}
\right. \\
\delta^{2} h^{\prime (n)}_{\left( \sigma, \Lambda_{j} \right)} &=& \left\{
\begin{array}{lll}
1-\frac{1}{n}  & \text{ for } &  j=0 \\
-1-\frac{1}{n}  & \text{ for } &  j=n-j_{-} \\
-\frac{1}{n}   & \text{ for } &  j \neq 0,n-j_{-}
\end{array}
\right.
\end{eqnarray}
so that the peaks are bunched into two groups of $n/2$ peaks, with one separation that is bunched closer than the rest within the first group, between the first $j_{+}+1$ and next $j_{-}$ peaks. We need to examine the cases $n=2$ and $4$ separately, since they have $j_{-}=0$.

For $n=2$, the tunneling and relaxation sequence is
\begin{eqnarray}
\ldots \xrightarrow{\text{relax}} \left( I, \Lambda_{0} \right) &\xrightarrow{e}& \left( \psi, \Lambda_{1} \right) \xrightarrow{\text{relax}} \left( \sigma, \Lambda_{0} \right) \notag \\
&\xrightarrow{e}& \left( \sigma, \Lambda_{1} \right) \xrightarrow{\text{relax}} \left( I, \Lambda_{0} \right) \xrightarrow{e} \ldots
\end{eqnarray}
which gives alternation between
\begin{eqnarray}
\delta^{2} h^{\prime (n)}_{\left( I, \Lambda_{0} \right)} &=& \frac{1}{2} \\
\delta^{2} h^{\prime (n)}_{\left( \sigma, \Lambda_{0} \right)} &=& -\frac{1}{2}
.
\end{eqnarray}

For $n=4$, the tunneling and relaxation sequence is
\begin{eqnarray}
\ldots &\xrightarrow{\text{relax}}& \left( I, \Lambda_{0} \right) \xrightarrow{e} \left( \psi, \Lambda_{1} \right) \xrightarrow{\text{relax}} \left( I, \Lambda_{1} \right) \notag \\
&\xrightarrow{e}& \left( \psi, \Lambda_{2} \right) \xrightarrow{\text{relax}} \left( \sigma, \Lambda_{0} \right) \xrightarrow{e} \left( \sigma, \Lambda_{1} \right) \notag \\
&\xrightarrow{\text{relax}}& \left( I, \Lambda_{3} \right) \xrightarrow{e} \left( \psi, \Lambda_{0} \right) \xrightarrow{\text{relax}} \left( I, \Lambda_{0} \right) \xrightarrow{e} \ldots
\end{eqnarray}
which gives
\begin{eqnarray}
\delta^{2} h^{\prime (n)}_{\left( I, \Lambda_{0} \right)} &=& \frac{3}{4} \\
\delta^{2} h^{\prime (n)}_{\left( I, \Lambda_{1} \right)} = \delta^{2} h^{\prime (n)}_{\left( \sigma, \Lambda_{0} \right)} = \delta^{2} h^{\prime (n)}_{\left( I, \Lambda_{3} \right)} &=& -\frac{1}{4}
,
\end{eqnarray}
i.e. the peaks bunch into groups of four.

The BS states built on the $\nu=1/2$ MR parent state have counterparts at the same filling fraction built instead on the $\nu=1/2$ $\overline{\text{Pf}}$ state (we denote these $\overline{\text{BS}}^{\psi}$). These are constructed by condensing Laughlin type quasiparticles in the $\overline{\text{Pf}}$ state and similarly possess a subset which admits a CF type description~\footnote{This is a slightly different construction from just particle-hole conjugating the BS$^{\psi}$ states constructed in Ref.~\onlinecite{Bonderson07d}, however they produce the same braiding statistics, edge modes, and conformal dimensions, so they are essentially the same for almost all purposes.}. Their CF type ground-state wavefunctions are given by
\begin{eqnarray}
\Psi_{\frac{n}{n+1}}^{( \overline{\text{BS}}^{\psi} )} &=& \mathcal{P}_{\text{LLL}} \left\{ \Psi_{1/2}^{(\overline{\text{Pf}})} \chi_{1}^{-1} \chi_{n}  \right\} \\
\Psi_{\frac{n}{3n-1}}^{ ( \overline{\text{BS}}^{\psi} )} &=& \mathcal{P}_{\text{LLL}} \left\{ \Psi_{1/2}^{( \overline{\text{Pf}} )} \chi_{1} \chi_{-n} \right\} \simeq \frac{\Psi_{1/2}^{( \overline{\text{Pf}} )} \Psi_{\frac{n}{2n-1}}^{\left( \text{CF} \right)} }{\chi_1}
\end{eqnarray}
The $\nu =\frac{n}{3n-1}$ series of these $\overline{\text{BS}}^{\psi}$ states have edge theories which can be described by $\overline{\text{SU}(2)}_2 \times \overline{\text{SU}(n)}_{1} \times \text{U}(1)$, and so exhibit exactly the same bunching patterns as the BS-CF states.

There are a few other BS states not included in the above analyses which are also of interest because they correspond to observed second Landau level quantum Hall states. We provide the Coulomb blockade peak spacings for these here, without the calculational details:

For the BS$_{1/3}^{\psi}$ and $\overline{\text{BS}}_{2/3}$ states at $\nu=1/3$, the pattern is given by $\delta^{2} h^{(n)} = 0$ for all bulk configurations of the puddle. The spacing pattern is the same when there is relaxation.

For the $\overline{\text{BS}}_{1/3}^{\psi}$ state at $\nu=2/3$, when the puddle contains an even number of $\sigma$-type quasiparticles, the spacing will alternate between $\delta^{2} h^{(n)} = \pm \frac{1}{2}$. When the puddle contains an odd number of $\sigma$-type quasiparticles, the spacing will be $\delta^{2} h^{(n)} = 0$. When there is relaxation, the spacing pattern will be given by $\delta^{2} h^{\prime (n)} = \pm \frac{1}{4}$ for all bulk configurations.

\subsection{Read-Rezayi States}
\label{sec:RR}

The $k$-clustered Read-Rezayi (RR$_{k,M}$) states~\cite{Read99} at $\nu= \frac{k}{kM+2}$ can be written as $\text{Pf}_k \times \text{U}(1)$, where the U$(1)$ is the charge sector and the $\mathbb{Z}_k$ parafermions~\cite{Zamolodchikov85,Gepner87} ($\text{Pf}_k$) is the neutral sector. The $\text{Pf}_k$ charges $\Phi_{m}^{2j}$ carry a SU$(2)_k$ charge $j$ and a $\mathbb{Z}_{2k}$ charge $m$, the pair of which are restricted to obey $\left[ 2j+m \right]_2 = 0$ and the identifications $\Phi_{m}^{2j} = \Phi_{m+2k}^{2j} = \Phi_{m\pm k}^{k-2j}$. Consequently, their fusion rules are given by
\begin{equation}
\Phi_{m_1}^{2j_1} \times \Phi_{m_2}^{2j_2} = \sum_{j = \left| j_1 -j_2 \right|}^{\text{min}\left\{ j_1 + j_2 , k - j_1 -j_2 \right\}} \Phi_{m_1 + m_2}^{2j}
,
\end{equation}
and their conformal dimensions are
\begin{equation}
\label{eq:h_Pf}
h_{\Phi_{m}^{2j}} = \left\{
\begin{array}{lll}
\frac{j\left(j+1\right)}{k+2} - \frac{m^2}{4k}  & \text{ for } &  \left| m \right| \leq 2j \\
\frac{j\left(j+1\right)}{k+2} - \frac{m^2}{4k} -j + \frac{ \left|m\right| }{2}  & \text{ for } &  2j \leq \left| m \right| \leq k
\end{array}
\right.
.
\end{equation}
Noting that the electron carries parafermion charge $\psi_1 = \Phi_{2}^{0}$, electron tunneling will give the sequence of parafermionic charges
\begin{equation}
\ldots \rightarrow \Phi_{m}^{2j} \rightarrow \Phi_{m+2}^{2j} \rightarrow \ldots
,
\end{equation}
for which one obtains
\begin{equation}
\delta^{2} h^{(n)}_{\Phi_{m}^{2j}} = \left\{
\begin{array}{lll}
2-\frac{2}{k}  & \text{ for } &  \left[ m \right]_{2k} = 2j = 0 \\
1-\frac{2}{k}  & \text{ for } &  \left[ m \right]_{2k} = \pm 2j \neq 0 \\
-\frac{2}{k}   & \text{ for } &  \left[ m \right]_{2k} \neq 2j \\
\end{array}
\right.
.
\end{equation}
This gives bunching of the resonance peaks into groups of $2j$ and $k-2j$. This matches the results of Ref.~\onlinecite{Ilan08a}

These states allow relaxation of the SU$(2)_k$ charge by tunneling neutral excitations carrying $\varepsilon_{j} = \Phi_{0}^{2j}$ charges (where $j$ is an integer) between the edge and bulk quasiparticles. This relaxes the edge to the minimal weight charges $\sigma_{m} = \Phi_{m}^{m}$ (note that $\Phi_{k}^{k}=\Phi_{0}^{0} = I$ and $\Phi_{k-1}^{k+1}=\Phi_{1}^{1} = \sigma_1$). Adding an electron to the puddle and then relaxing the system in this manner, one has the tunneling and relaxation sequence
\begin{equation}
\ldots \xrightarrow{\text{relax}} \Phi_{m}^{m}  \xrightarrow{e} \Phi_{m+2}^{m} \xrightarrow{\text{relax}} \Phi_{m+2}^{m+2} \xrightarrow{e} \ldots
\end{equation}
with
\begin{equation}
\delta^{2} h^{\prime (n)}_{\Phi_{m}^{m}} = \left\{
\begin{array}{lll}
1-\frac{2}{k}  & \text{ for } &  m = 0, 1 \\
-\frac{2}{k}   & \text{ for } &  m \neq 0,1 \\
\end{array}
\right.
.
\end{equation}
When $k$ is even, this gives bunching of the resonance peaks into groups of $k/2$. When $k$ is odd, this gives bunching of the peaks into alternating groups of $\frac{k-1}{2}$ and $\frac{k+1}{2}$. This matches the results of Ref.~\onlinecite{Ilan08b}.

As previously mentioned, the Coulomb blockade patterns (both with and without relaxation) of the RR$_{k,M}$ states are identical to those found for the Abelian $k$-component $[M+2;M]$ states and their hierarchical counterparts in Sec.~\ref{sec:kcomponent}.

\subsection{Anti-Read-Rezayi States}
\label{sec:ARR}

Particle-hole conjugating the RR$_k$ states ($M=1$) and assuming the neutral edge mode equilibrate, one has the $\overline{\text{RR}}_{k}$ states at $\nu = \frac{2}{k+2}$ with edge theory described by $\overline{\text{SU}(2)}_{k} \times \text{U}(1)$, where the $\text{U}(1)$ is the charge sector and $\overline{\text{SU}(2)}_{k}$ is the neutral sector~\cite{Bishara08b}. The electrons carry $\overline{\text{SU}(2)}_{k}$ charge $\frac{k}{2}$. The fusion rules for $\text{SU}(2)_{k}$ are given by
\begin{equation}
j_1 \times j_2 = \sum_{j = \left| j_1 -j_2 \right|}^{\text{min}\left\{ j_1 + j_2 , k - j_1 -j_2 \right\}  } j
\end{equation}
(in particular, $j \times \frac{k}{2} = \frac{k}{2} -j$) and the conformal dimensions are
\begin{equation}
h_{j} = \frac{j \left( j+1 \right)}{k+2}
.
\end{equation}
Thus, when $j$ is the total $\overline{\text{SU}(2)}_{k}$ charge of the bulk quasiparticles in the puddle, electron tunneling will give the sequence
\begin{equation}
\ldots \rightarrow j \rightarrow \frac{k}{2} - j \rightarrow j \rightarrow \ldots
,
\end{equation}
for which one has alternation between
\begin{eqnarray}
\delta^{2} h^{(n)}_{j} &=& \frac{k-4j}{2} ,\\
\delta^{2} h^{(n)}_{\frac{k}{2}-j} &=& -\frac{k-4j}{2}
.
\end{eqnarray}
This gives bunching of the resonance peaks into pairs. This bunching pattern is identical to those of states described in Sec.~\ref{sec:BSARR} (with $k$ here matching up with $k-1$ there).

These states allow relaxation of the $\overline{\text{SU}(2)}_{k}$ charge by tunneling neutral excitations carrying integer $\overline{\text{SU}(2)}_{k}$ charges between the edge and bulk quasiparticles. This relaxes the edge to either $0$ or $\frac{1}{2}$ $\overline{\text{SU}(2)}_{k}$ charge (depending on whether $j$ was an integer or half-integer). For $k$ even, this gives either of the following two tunneling and relaxation sequences
\begin{eqnarray}
&& \ldots \xrightarrow{\text{relax}} 0 \xrightarrow{e} \frac{k}{2} \xrightarrow{\text{relax}} 0 \xrightarrow{e} \ldots \\
&& \ldots \xrightarrow{\text{relax}} \frac{1}{2} \xrightarrow{e} \frac{k-1}{2} \xrightarrow{\text{relax}} \frac{1}{2} \xrightarrow{e} \ldots
\end{eqnarray}
which both have
\begin{equation}
\delta^{2} h^{\prime (n)}_{0} =\delta^{2} h^{\prime (n)}_{\frac{1}{2}} = 0
.
\end{equation}
For $k$ odd, the tunneling and relaxation sequence will be
\begin{equation}
\ldots \xrightarrow{\text{relax}} 0 \xrightarrow{e} \frac{k}{2} \xrightarrow{\text{relax}} \frac{1}{2} \xrightarrow{e} \frac{k-1}{2} \xrightarrow{\text{relax}} 0 \xrightarrow{e} \ldots
\end{equation}
for which there is alternation between
\begin{eqnarray}
\delta^{2} h^{\prime (n)}_{0} &=& \frac{1}{2} ,\\
\delta^{2} h^{\prime (n)}_{\frac{1}{2}} &=& -\frac{1}{2}
.
\end{eqnarray}

We note that the SU$(2)_k$ NAF states~\cite{Blok92} (which include filling fractions $\nu=\frac{2}{k+2}$ for $k$ even and $\nu=\frac{2}{k+4}$ for $k$ odd) have edge theories given by $\text{SU}(2)_{k} \times \text{U}(1)$, where the U$(1)$ is purely the charge sector, so the resonance peak patterns found for $\overline{\text{RR}}_{k}$ also apply to the SU$(2)_k$ NAF states.

\subsection{Hierarchy States over anti-Read-Rezayi}
\label{sec:BSARR}

One can apply the BS hierarchy construction~\cite{Bonderson07d} to the $\overline{\text{RR}}_{k}$ state at $\nu=\frac{2}{k+2}$. Building the hierarchy in the charge sector by condensing a gas of charge $\frac{2e}{k+2}$ Laughlin quasiholes in the first step, produces states described by $\overline{\text{Pf}}_k \times \text{U}(1)_K$. The corresponding CF type ground-state wavefunctions for these are
\begin{equation}
\Psi_{\frac{2n}{kn+4n-2}}^{ ( \text{BS-}\overline{\text{RR}} )} = \mathcal{P}_{\text{LLL}} \left\{ \Psi_{\frac{2}{k+2}}^{( \overline{\text{RR}} )} \chi_{1} \chi_{-n} \right\} \simeq \frac{\Psi_{\frac{2}{k+2}}^{( \overline{\text{RR}} )} \Psi_{\frac{n}{2n-1}}^{\left( \text{CF} \right)} }{\chi_1}
\end{equation}
The edge theory for such states can be described by $\overline{\text{SU}(2)}_{k} \times \overline{\text{SU}(n)}_{1} \times \text{U}(1)$, where the $\text{U}(1)$ is the charge sector and $\overline{\text{SU}(2)}_{k} \times \overline{\text{SU}(n)}_{1}$ is the neutral sector. The electrons carry $\overline{\text{SU}(2)}_{k} \times \overline{\text{SU}(n)}_{1}$ charge $\left( \frac{k}{2} , \Lambda_1 \right)$.

We now restrict our attention to the states at the first level of hierarchy ($n=2$). These have filling $\nu=\frac{2}{k+3}$ and $K$-matrix
\begin{equation}
K = \left[
\begin{array}{rrr}
1     &     1             &     1     \\
1     &   -\frac{2}{k}    &     0     \\
1     &      0            &    -2
\end{array}
\right]
.
\end{equation}
When the total $\overline{\text{SU}(2)}_{k} \times \overline{\text{SU}(2)}_{1}$ topological charge of the bulk quasiparticles in the puddle is $\left( j, \Lambda_{a} \right)$, electron tunneling will give the sequence
\begin{equation}
\ldots \rightarrow \left( j, \Lambda_{a} \right) \rightarrow \left( \frac{k}{2} - j , \Lambda_{\left[a+1\right]_2} \right) \rightarrow \left( j, \Lambda_a \right) \rightarrow \ldots
,
\end{equation}
for which one has alternation between
\begin{eqnarray}
\delta^{2} h^{(n)}_{\left( j, \Lambda_a \right)} &=& \frac{k-4j+\left(-1\right)^{a}}{2} ,\\
\delta^{2} h^{(n)}_{\left(\frac{k}{2}-j , \Lambda_{\left[a+1\right]_2} \right)} &=& -\frac{k-4j+\left(-1\right)^{a}}{2}
.
\end{eqnarray}
This gives bunching of the resonance peaks into pairs. These bunching patterns are identical to those produced by the $\overline{\text{RR}}_{k+1}$ states at $\nu=\frac{2}{k+3}$ (see Sec.~\ref{sec:ARR}).

These states allow relaxation of the $\overline{\text{SU}(2)}_{k} \times \overline{\text{SU}(2)}_{1}$ charge by tunneling neutral excitations carrying $\left( j, \Lambda_{a} \right)$ where $\left[ 2j+a \right]_2 =0$ between the edge and bulk quasiparticles. This relaxes the edge to either $0$ or $\frac{1}{2}$ $\overline{\text{SU}(2)}_{k}$ charge (depending on $j$, $a$, and $k$).

For $k$ odd, this gives either of the following two tunneling and relaxation sequences
\begin{equation}
\ldots \xrightarrow{\text{relax}} \left( 0, \Lambda_{0} \right) \xrightarrow{e} \left( \frac{k}{2}, \Lambda_{1} \right) \xrightarrow{\text{relax}} \left( 0, \Lambda_{0} \right) \xrightarrow{e} \ldots
\end{equation}
\begin{equation}
\ldots \xrightarrow{\text{relax}} \left( \frac{1}{2}, \Lambda_{0} \right) \xrightarrow{e} \left( \frac{k-1}{2}, \Lambda_{1} \right) \xrightarrow{\text{relax}} \left( \frac{1}{2}, \Lambda_{0} \right) \xrightarrow{e} \ldots
\end{equation}
which both have
\begin{equation}
\delta^{2} h^{\prime (n)}_{0} =\delta^{2} h^{\prime (n)}_{\frac{1}{2}} = 0
.
\end{equation}

For $k$ even, the tunneling and relaxation sequence will be
\begin{eqnarray}
\ldots \xrightarrow{\text{relax}} \left( 0, \Lambda_{0} \right) \xrightarrow{e} \left( \frac{k}{2}, \Lambda_{1} \right) &\xrightarrow{\text{relax}}& \left( \frac{1}{2}, \Lambda_{0} \right) \notag \\
\xrightarrow{e} \left( \frac{k-1}{2}, \Lambda_{1} \right) &\xrightarrow{\text{relax}}& \left( 0, \Lambda_{0} \right) \xrightarrow{e} \ldots
\end{eqnarray}
for which there is alternation between
\begin{eqnarray}
\delta^{2} h^{\prime (n)}_{0} &=& \frac{1}{2} ,\\
\delta^{2} h^{\prime (n)}_{\frac{1}{2}} &=& -\frac{1}{2}
.
\end{eqnarray}
These bunching patterns are identical to those produced by the $\overline{\text{RR}}_{k+1}$ states at $\nu=\frac{2}{k+3}$ (see Sec.~\ref{sec:ARR}).

Thus, these (first level) hierarchy states at $\nu=\frac{2}{k+3}$ built over the $\overline{\text{RR}}_{k}$ are Coulomb blockade doppelg\"{a}ngers of the $\overline{\text{RR}}_{k+1}$ states at $\nu=\frac{2}{k+3}$, both with and without relaxation. (We also note that both these $\nu=\frac{2}{k+3}$ states have shift $S=-k$ on the sphere.)

\subsection{Second Landau Level States}

The lowest Landau level quantum Hall states are all strongly expected to be Abelian states described by the Laughlin and HH states. On the other hand, the physics of the second Landau level is far less certain, but strongly expected to possess non-Abelian topological orders. Well-developed quantum Hall states have been observed in the second Landau level at filling fractions $\nu=5/2$, $7/3$, $8/3$, $14/5$, and $12/5$~\cite{Pan99,Xia04}. We now focus on the relevant candidate quantum Hall states for these filling fractions (neglecting $\nu=14/5$, which is strongly expected to simply be an Abelian particle-hole conjugate Laughlin state).

\subsubsection{$\nu=5/2$}

The $\nu=5/2$ candidates MR (RR$_2$), $\overline{\text{Pf}}$ ($\overline{\text{RR}}_{2}$), SU$(2)_2$ NAF, $(3,3,1)$ ($2$-component $[3;1]$), its hierarchical counterpart, and BS-$\overline{\text{L}}_{1/3}$ (BS-$\overline{\text{RR}}_{1}$) all have identical Coulomb blockade patterns. Specifically, for an even number of fundamental quasiparticles in the bulk, the spacing between Coulomb blockade resonance peaks will alternate between
\begin{equation}
\Delta A_{0} = \frac{e}{\rho_0} \left( 1 \pm \frac{ v_n }{ 2 v_c } \right)
.
\end{equation}
For an odd number of fundamental quasiparticles in the bulk, one will simply have
\begin{equation}
\Delta A_{1} = \frac{e}{\rho_0}
.
\end{equation}

When relaxation occurs, all of these states will always have the trivial spacing pattern $\Delta A = \frac{e}{\rho_0}$ between peaks.

\subsubsection{$\nu=7/3$}

The $\nu=7/3$ candidates Laughlin (L$_{1/3}$), BS$_{1/3}^{\psi}$, and $\overline{\text{BS}}_{2/3}$ all have identical Coulomb blockade patterns. Specifically, these always exhibit the spacing
\begin{equation}
\Delta A = \frac{e}{\rho_0}
\end{equation}
between peaks.

The $\overline{\text{RR}}_{4}$ and BS-$\overline{\text{RR}}_{3}$ states will exhibit spacing patterns that alternate between
\begin{equation}
\Delta A_{j} = \frac{e}{\rho_0} \left( 1 \pm \frac{ 4 (j-1) v_n }{ 3 v_c } \right)
\end{equation}
where $j=0$, $\frac{1}{2}$, $1$, $\frac{3}{2}$, or $2$, depending on the bulk quasiparticle configuration.

When relaxation occurs, L$_{1/3}$, BS$_{1/3}^{\psi}$, $\overline{\text{BS}}_{2/3}$, $\overline{\text{RR}}_{4}$, and BS-$\overline{\text{RR}}_{3}$ all have the same trivial spacing pattern $\Delta A = \frac{e}{\rho_0}$. We note that the L$_{1/3}$, BS$_{1/3}^{\psi}$, and $\overline{\text{BS}}_{2/3}$ states will also be quite difficult to distinguish from each other using tunneling and interferometry experiments~\cite{Bishara09}. It seems that thermal transport experiments may be the best hope for distinguishing between these.

\subsubsection{$\nu=8/3$}

The $\nu=8/3$ candidates have distinct Coulomb blockade patterns. The $\overline{\text{L}}_{1/3}$ (HH$_{2/3}$) state will exhibit spacing between resonance peaks that alternate between
\begin{equation}
\Delta A = \frac{e}{\rho_0} \left( 1 \pm \frac{ v_n }{ 3 v_c } \right)
.
\end{equation}

The BS$_{2/3}$ state will exhibit two possible spacing patterns, depending on the bulk quasiparticle configuration: alternation between
\begin{equation}
\Delta A_{0} = \frac{e}{\rho_0} \left( 1 \pm \frac{ v_n }{ v_c } \right)
\end{equation}
or alternation between
\begin{equation}
\Delta A_{1} = \frac{e}{\rho_0} \left( 1 \pm \frac{ v_n }{ 3 v_c } \right)
.
\end{equation}

The $\overline{\text{BS}}_{1/3}^{\psi}$ state will exhibit two possible spacing patterns, depending on the bulk quasiparticle configuration: alternation between
\begin{equation}
\Delta A_{0} = \frac{e}{\rho_0} \left( 1 \pm \frac{ v_n }{ 3 v_c } \right)
\end{equation}
or simply the trivial spacing
\begin{equation}
\Delta A_{1} = \frac{e}{\rho_0}
.
\end{equation}

The RR$_{4}$ state will exhibit three possible spacing patterns, depending on the bulk quasiparticle configuration: bunching into groups of four, alternating bunching into groups of three and one, and bunching into groups of two. The bunched spacing within a group is always
\begin{equation}
\Delta A = \frac{e}{\rho_0} \left( 1 - \frac{ v_n }{ 3 v_c } \right)
.
\end{equation}
The spacing between consecutive bunched groups of four is
\begin{equation}
\Delta A = \frac{e}{\rho_0} \left( 1 + \frac{ v_n }{ v_c } \right)
,
\end{equation}
while the spacing between consecutive bunched groups is otherwise
\begin{equation}
\Delta A = \frac{e}{\rho_0} \left( 1 + \frac{ v_n }{ 3 v_c } \right)
.
\end{equation}

When relaxation occurs, $\overline{\text{L}}_{1/3}$, BS$_{2/3}$, and RR$_{4}$ all have identical Coulomb blockade patterns, exhibiting alternation between
\begin{equation}
\Delta A = \frac{e}{\rho_0} \left( 1 \pm \frac{ v_n }{ 3 v_c } \right)
.
\end{equation}
With relaxation, $\overline{\text{BS}}_{1/3}^{\psi}$ only exhibits alternation between
\begin{equation}
\Delta A = \frac{e}{\rho_0} \left( 1 \pm \frac{ v_n }{ 6 v_c } \right)
.
\end{equation}

\subsubsection{$\nu=12/5$}

The $\nu=12/5$ candidates BS$_{2/5}$, $\overline{\text{BS}}_{3/5}^{\psi}$ (BS-$\overline{\text{RR}}_{2}$), and $\overline{\text{RR}}_{3}$ have identical Coulomb blockade patterns. Specifically, depending on the bulk quasiparticle configuration, they will exhibit two possible spacing patterns: alternation between
\begin{equation}
\Delta A_{0} = \frac{e}{\rho_0} \left( 1 \pm \frac{ 3 v_n }{ 5 v_c } \right)
\end{equation}
or alternation between
\begin{equation}
\Delta A_{1} = \frac{e}{\rho_0} \left( 1 \pm \frac{ v_n }{ 5 v_c } \right)
.
\end{equation}

When relaxation occurs, these states all have the same Coulomb blockade patterns exhibited by the HH$_{2/5}$ state (with or without relaxation), which is alternation between
\begin{equation}
\Delta A = \frac{e}{\rho_0} \left( 1 \pm \frac{ v_n }{ 5 v_c } \right)
.
\end{equation}

\section{Conclusion}

We have demonstrated that Coulomb blockade experiments, while possibly somewhat useful in rather limited contexts, are generally quite poor at distinguishing and identifying topological orders, particularly for the purposes of the quantum Hall states expected to be non-Abelian. This re-emphasizes the value of interference experiments, which are capable of directly probing quasiparticle braiding statistics~\cite{Chamon97,Fradkin98,DasSarma05,Stern06a,Bonderson06a,Bonderson06b,Bonderson07c,Bishara09}, and thus offer the best method of identifying the topological order of a system. It may also be useful to supplement interference experiments with ones that measure the scaling properties and/or thermal transport, which could potentially provide extra details that interferometry might miss, such as quasiparticle scaling dimensions~\cite{Wen92b} and the chiral central charge. However, such experiments depend crucially on details of the edge physics which may be prone to non-universal effects that debase the information gained from them.

\begin{acknowledgments}
We thank S.~Simon for useful discussions. We acknowledge the hospitality of the Aspen Center for Physics. KS acknowledges the support and hospitality of Microsoft Station Q. CN and KS are supported in part by the DARPA-QuEST program. KS is supported in part by the NSF under grant DMR-0748925.
\end{acknowledgments}


\end{document}